\documentclass[aps, twocolumn, prx,superscriptaddress, longbibliography, raggedbottom]{revtex4-2}
\usepackage{graphicx}
\usepackage{dcolumn}
\usepackage{bm}
\usepackage{amsmath}
\usepackage[dvipsnames]{xcolor}
\usepackage[utf8]{inputenc}

\begin{document}

\preprint{APS/123-QED}

\title{Hund’s coupling assisted orbital-selective superconductivity in Ba$_{1-x}$K$_{x}$Fe$_{2}$As$_{2}$}

\author{Elena Corbae}
\affiliation{Stanford Institute for Materials and Energy Sciences, SLAC National Accelerator Laboratory, 2575 Sand Hill Road, Menlo Park, California 94025, USA}
\affiliation{Geballe Laboratory for Advanced Materials, Department of Applied Physics and Physics, Stanford University, Stanford, California 94305, USA}
\affiliation{Department of Applied Physics, Stanford University, Stanford, California 94305, USA}
\thanks{To whom correspondence should be addressed; Email: ecorbae@stanford.edu; zxshen@stanford.edu}
\author{Rong Zhang}
\affiliation{Stanford Institute for Materials and Energy Sciences, SLAC National Accelerator Laboratory, 2575 Sand Hill Road, Menlo Park, California 94025, USA}
\affiliation{Department of Applied Physics, Stanford University, Stanford, California 94305, USA}
\author{Cong Li}
\affiliation{Department of Applied Physics, KTH-Royal
Institute of Technology, Stockholm 11419, Sweden}
\author{Kunihiro Kihou}
\affiliation{National Institute of Advanced Industrial Science and Technology (AIST), Tsukuba, Japan}
\author{Chul-Ho Lee}
\affiliation{National Institute of Advanced Industrial Science and Technology (AIST), Tsukuba, Japan}
\author{Makoto Hashimoto}
\affiliation{Stanford Synchrotron Radiation Lightsource, SLAC National Accelerator Laboratory, 2575 Sand Hill Road, Menlo Park, California 94025, USA}
\author{Thomas Devereaux}
\affiliation{Stanford Institute for Materials and Energy Sciences, SLAC National Accelerator Laboratory, 2575 Sand Hill Road, Menlo Park, California 94025, USA}
\affiliation{Department of Material Science, Stanford University, Stanford, California 94305, USA}
\author{Oscar Tjernberg}
\affiliation{Department of Applied Physics, KTH-Royal
Institute of Technology, Stockholm 11419, Sweden}
\author{Egor Babaev}
\affiliation{Department of Physics, KTH-Royal Institute of Technology, SE-10691 Stockholm, Sweden}
\affiliation{Wallenberg Initiative Materials Science for Sustainability, Department of Physics, KTH Royal Institute of Technology, SE-106 91 Stockholm, Sweden}
\author{Dung-Hai Lee}
\affiliation{Department of Physics, University of California, Berkeley, CA 94720, USA}
\affiliation{Material Sciences Division, Lawrence Berkeley National Laboratory, Berkeley, CA 94720, USA}
\author{Vadim Grinenko}
\affiliation{Tsung-Dao Lee Institute and School of Physics and Astronomy, Shanghai Jiao Tong University, Shanghai 201210, China}
\author{Donghui Lu}
\affiliation{Stanford Synchrotron Radiation Lightsource, SLAC National Accelerator Laboratory, 2575 Sand Hill Road, Menlo Park, California 94025, USA}
\author{Zhi-Xun Shen}
\affiliation{Stanford Institute for Materials and Energy Sciences, SLAC National Accelerator Laboratory, 2575 Sand Hill Road, Menlo Park, California 94025, USA}
\affiliation{Geballe Laboratory for Advanced Materials, Department of Applied Physics and Physics, Stanford University, Stanford, California 94305, USA}
\affiliation{Department of Applied Physics, Stanford University, Stanford, California 94305, USA}
\affiliation{Department of Physics, Stanford University, Stanford, California 94305, USA}

\date{\today}

\begin{abstract}
While the superconducting transition temperature of hole-doped Ba$_{1-x}$K$_{x}$Fe$_{2}$As$_{2}$ decreases past optimal doping, superconductivity does not completely disappear even for the fully doped KFe$_{2}$As$_{2}$ compound. In fact, superconductivity is robust through a Lifshitz transition where electron bands become hole-like around the zone corner at $x\sim0.7$, thus challenging the conventional understanding of superconductivity in iron-based systems. High resolution angle-resolved photoemission spectroscopy is used to investigate the superconducting gap structure, as well as the normal state electronic structure,  around optimal doping and across the Lifshitz transition. Our findings reveal a largely orbital-dependent superconducting gap structure, where the more strongly correlated $d_{xy}$ band has a vanishing superconducting gap at higher doping, aligning with the Hund’s metal behavior observed in the normal state. Notably, the superconducting gap on the $d_{xy}$ band disappears before the Lifshitz transition, suggesting that the Fermi surface topology may play a secondary role. We discuss how these results point to orbital-selective superconducting pairing and how strong correlations via Hund’s coupling may shape superconducting gap structures in iron-based and other multiorbital superconductors.
\end{abstract}

\maketitle

\section{\label{sec:level1}Introduction}

A critical problem in formulating the theory of high temperature superconductivity is identifying the pairing symmetry of the superconducting gap, which is intimately related to the underlying pairing mechanism. For iron-based superconductors, distinct pairing symmetries have been proposed based on inconsistent experimental results \cite{wangtext, Fernandes2022, Bang_2017, spm_PhysRevLett.101.057003, orbfluctuationPhysRevB.78.195114, octetlinenodedoi:10.1126/science.1222793, Xu2011, unconventionalpairing111PhysRevLett.101.087004, Hirschfeld_2011}, thus making it difficult to determine the pairing glue. As a model compound of the iron-based superconductors, the phase diagram of Ba$_{1-x}$K$_{x}$Fe$_{2}$As$_{2}$  shows that the superconducting phase arises out of a magnetically ordered parent phase. This supports a leading theoretical proposal of superconductivity in the iron-based systems - spin fluctuation mediated pairing with $s_\pm$ symmetry. In this model, the antiferromagnetic spin fluctuations create an effective attraction across Fermi surfaces (FSs) that peaks near the nesting vector connecting hole pockets at the zone center and electron pockets at the zone corner, thus relying on the condition that the electron and hole pockets are well nested \cite{spm_PhysRevLett.101.057003, unconventionalpairing111PhysRevLett.101.087004, Hirschfeld_2011, fsnestingexperimental}.\par
The importance of favorable nesting Fermiology for superconductivity is supported by experiments on the electron-doped system $\text{Ba}{({\text{Fe}}_{1\ensuremath{-}x}{\text{Co}}_{x})}_{2}{\text{As}}_{2}$ \cite{fsnestingexperimental, PhysRevB.82.054515}. Neutron scattering finds the spin resonance decreases with the increasing mismatch of the hole and electron FSs until it eventually goes away, corresponding with the disappearance of superconductivity \cite{PhysRevB.82.054515}. However, the hole over-doped Ba$_{1-x}$K$_{x}$Fe$_{2}$As$_{2}$ presents as a complication for the nesting picture. Here, the system undergoes a Lifshitz transition with hole-doping as electron pockets around the zone corner disappear while four petal-like hole pockets emerge. This change of Fermiology is consistent with neutron measurements that see a weakening of the spin resonance with the increasing mismatch between hole and electron FSs, as well as a splitting of the well-defined peak at $Q$ into two incommensurate peaks beyond optimal doping and before the Lifshitz transition \cite{Lee2016, PhysRevLett.107.177003, PhysRevLett.124.017001}. At the same time, the hole-doped side remains superconducting for all potassium-doping levels, even across the Lifshitz transition. This departure from the nesting picture that emerged from the electron-doped side has led to speculation that the bands participating in pairing may change \cite{Grinenko2020, PhysRevLett.107.117001}. Furthermore, the presence of multiple FSs allows for various superconducting scattering channels, such as intra- and inter-band, intra- and inter-orbital, and pair scattering among different FS sheets, thus adding another layer of complexity to understanding the superconducting pairing mechanism \cite{Graser_2009, fsscatteringchannels, Hirschfeld_2011, orbitalnesting}.\par

This multiband nature of Ba$_{1-x}$K$_{x}$Fe$_{2}$As$_{2}$ has also been shown to support exotic superconducting phenomena. Breaking of  time reversal symmetry (BTRS), was reported in  Ba$_{1-x}$K$_{x}$Fe$_{2}$As$_{2}$ in a narrow range of doping, both below and slightly above the superconducting phase transition by a variety of probes \cite{Grinenko2020, grinenko2021state,shipulin2023calorimetric,halcrow2024probing}. Additionally, several experiments reported vortices carrying an arbitrary fraction of the flux quantum \cite{iguchidoi:10.1126/science.abp9979,zheng2024direct,zhou2024observation}. For both examples, theoretical proposals suggest that the multiband nature of Ba$_{1-x}$K$_{x}$Fe$_{2}$As$_{2}$ is responsible for the displayed behavior \cite{Grinenko2020, zheng2024directobservationquantumvortex, iguchidoi:10.1126/science.abp9979}. When considering multiband materials with nondegenerate orbitals, Hund's coupling must be taken into account. An important  consequence of Hund’s coupling is a reduction of inter-orbital fluctuations, such that correlations can affect each orbital in a distinct manner: orbitals that are closer to half filling localize first, while the other orbitals may remain itinerant \cite{mottparentphase, annurev:/content/journals/10.1146/annurev-conmatphys-020911-125045}. This has been observed experimentally with phenomena such as orbital selective Mott phases and orbital dependent coherence-incoherence crossovers \cite{PhysRevLett.110.067003, PhysRevB.96.201108, PhysRevB.94.201109} in the normal state.\par
Given that the multiband nature of Ba$_{1-x}$K$_{x}$Fe$_{2}$As$_{2}$ is at the heart of many open questions of this material system, we employed angle-resolved photoemission spectroscopy (ARPES) to study the orbital-selective phenomena.  Combining high resolution ARPES measurements and density functional theory (DFT) calculations, we study the evolution of the superconducting gap structure and normal state properties with hole-doping. We observe a distinctive orbital dependent gap structure and clear signatures of orbital selective correlations, and discuss how the increased influence of Hund's coupling with hole-doping may manifest in the superconducting gap structure. These results support the survival of superconductivity with hole-doping and draw similarities to other strongly correlated superconducting systems.

\begin{figure*}
\includegraphics[width=\textwidth]{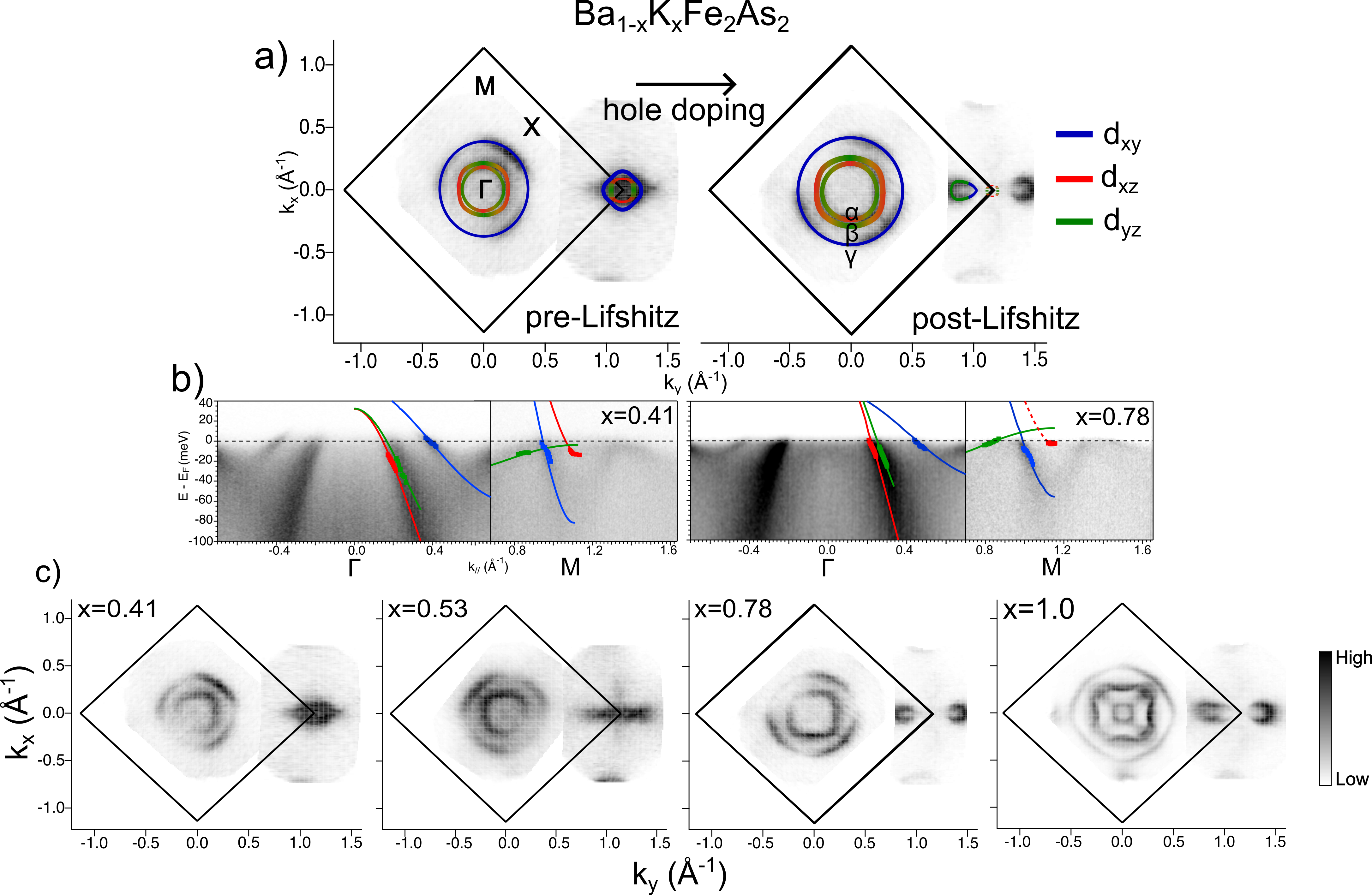}
\caption{\textbf{Ba$_{1-x}$K$_x$Fe$_2$As$_2$ Fermiology}. 
\textbf{(a)} Schematic of the Lishitz transition that takes place above optimal doping around the M point. Solid lines are fitted FS’, dotted line shows incipient FS. In the $\Gamma-X$ direction the orbital character is not well defined and thus is shown as a mix of red and green. Only two of the FSs around M are shown for simplicity, although the FS displays $C_4$ symmetry. \textbf{(b)} Spectrum of $x=0.41$ (left) and $x=0.78$ (right) in the $\Gamma-M$ direction, with renormalized DFT bands overlayed. 
\textbf{(c)} Experimental FS for dopings of $x=0.41$, $x=0.53$, $x=0.78$, $x=1.0$. The solid black line shows the Brillouin Zone.}
\label{fig:fermiologynesting}
\end{figure*}

\section{\label{sec:fermiology}Fermiology}
We study four doping levels of Ba$_{1-x}$K$_{x}$Fe$_{2}$As$_{2}$ ($x=0.41$, $x=0.53$, $x=0.78$, $x=1.0$) with a $T_{C}$ of 36.4 K, 25.8 K, 12 K and 5.5 K, respectively. We first identify the bands near the Fermi level in the normal state and confirm that a Lifshitz transition occurs between $x=0.53$ and $x=0.78$. The FS consists of three hole pockets ($\alpha, \beta, \gamma$) centered around $\Gamma$. The orbital character is shown in \figurename{\ref{fig:fermiologynesting}}(a), where in the $\Gamma-M$ direction the dominant orbital character of these three pockets are $d_{xz}$, $d_{yz}$ and $d_{xy}$, respectively. In the $\Gamma-X$ direction there is a strong mixture of $d_{xz}/d_{yz}$ character in the $\alpha$ and $\beta$ bands.  As shown in \figurename{\ref{fig:fermiologynesting}}(c), we indeed see that with hole doping the electron pockets around $M$ evolve into petal like hole pockets, consisting of bands of $d_{yz}/d_{xy}$ character. However, at a doping level of $x=0.78$ we still see signs of an electron-like band around $E_F$, and this feature only appears in the experimental geometry that the $d_{xz}$ electron band is seen \figurename{\ref{fig:mbandssuper}}(d-f). Around $M$, the $d_{yz}$ hole band has $C_4$ symmetry imposed degeneracy with the $d_{xz}$ electron band, and we confirm that at $x=0.78$ the $d_{yz}$ hole band crosses $E_F$, thus we do not expect to see the $d_{xz}$ electron band below $E_F$. There is not strong evidence from our measurements that $C_4$ symmetry is either broken or preserved. We note that a similar electron band has also been reported at a higher doping ($T_C$ = 9 K) by another study \cite{PhysRevB.88.220508}, where they assign it to be a trivial resolution tail of the $d_{xz}$ band. Our studies of KFe$_2$As$_2$ (n=5.5) also show this feature, see Supplemental Material (SM) Fig.S13 for details \cite{supp} (see also references \cite{crystalgrowthdoi:10.7566/JPSJ.85.034718, normanfunctionPhysRevB.57.R11093, phasediagrambak} therein). Dynamical Mean Field Theory (DMFT) of RbFe$_2$As$_2$ (n=5.5) shows that there may be some residual density of states around the Fermi level for the $d_{xz}$ orbital \cite{Chang2025}, thus we tentatively assign this electron-like band to have $d_{xz}$ character.

\section{\label{sec:superconductingmeasuremetns}Superconducting  Gap Measurements}
\begin{figure*}
    \includegraphics[width=\textwidth]{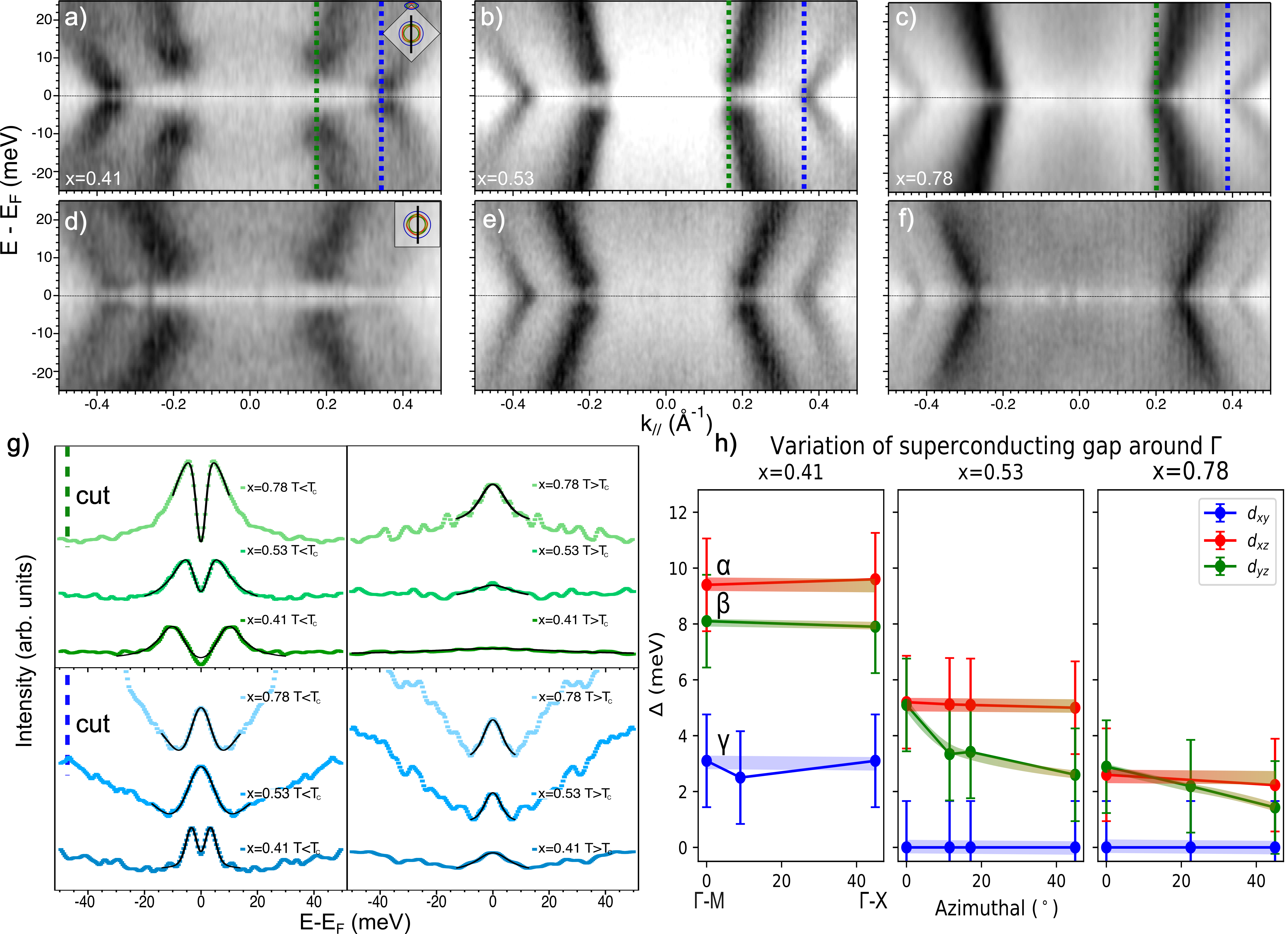}
    \caption{\textbf{Superconductivity around} $\boldsymbol{\Gamma}$. \textbf{(a)} Superconducting spectrum of $x=0.41$ in the $\Gamma-M$ direction. Green/blue dotted line shows where the Energy Distribution Curve (EDC) is taken for (g)/(h). Inset shows the experimental geometry for the spectrum cut with respect to the FS.\textbf{(b)-(c)} Same as (a) but for doping levels of $x=0.53$, $x=0.78$. \textbf{(d)} Schematic of cut direction on the FS for (f)-(h) spectrum. \textbf{(f)} Superconducting spectrum of $x=0.41$ in the $\Gamma-X$ direction. Inset shows the experimental geometry for the spectrum cut. \textbf{(e)-(f)} Same as (f) but for doping levels of x$=0.53$, $x=0.78$. \textbf{(g)} (left) EDCs at the at momentum indicated by the green and blue lines shown in (a)-(c) representing the $d_{yz}$(green) and $d_{xy}$(blue) bands, respectively. Dotted green/blue lines indicate EDCs for $T<T_C$, and solid lines are the superconducting gap fit. (right) Dotted lines show EDCs taken at $T>T_C$ at the same $k_F$, the solid lines overlaid are the fitted gap. There is no detecatable gap indicating the gap closes above $T_C$. The legend shows colors corresponding to doping levels of $x=0.41$, $x=0.53$, $x=0.78$. From the EDCs it is seen that the superconducting gap on the $d_{xy}$ band closes for doping levels of $x=0.53$ and above, while the gap persists on the $d_{yz}$ band for all doping levels. \textbf{(h)} Variation of the superconducting gap for all three doping levels as the azimuthal angle from $\Gamma-M$ towards $\Gamma-X$ is changed.}
    \label{fig:gammabandssuper}
\end{figure*}

\begin{figure*}
\includegraphics[width=\textwidth]{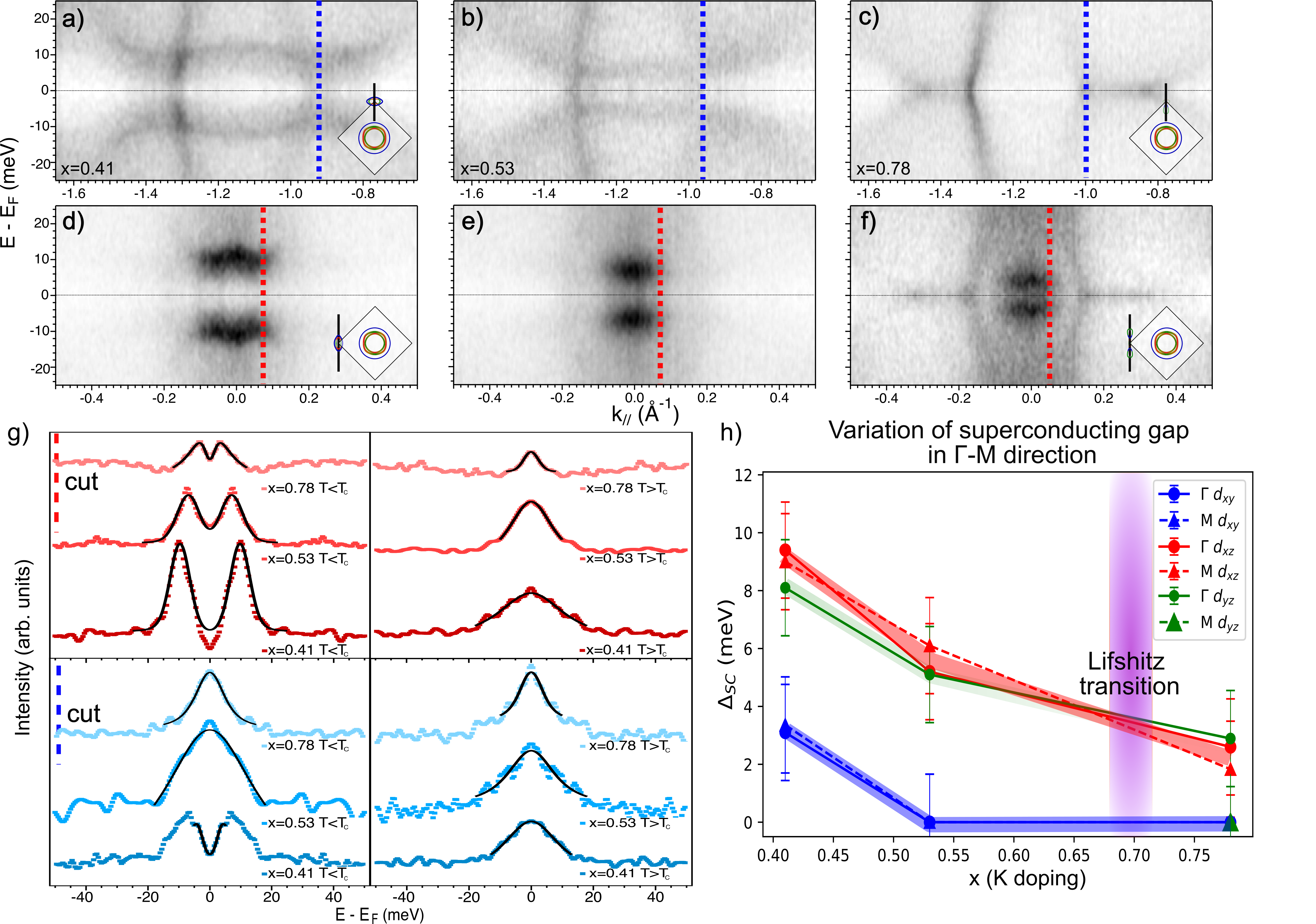}
\caption{\textbf{Superconductivity around $\boldsymbol{M}$.} 
\textbf{(a)} Spectrum of $x=0.41$  below $T_c$. Blue dotted line shows where the EDC is taken for (i). Inset schematic shows where the spectrum cut is taken with respect to the FS in an experimental geometry to maximize the signal of the $d_{xy}$ orbital around $M$. 
\textbf{(b)} Same as (a) but for $x=0.53$.
\textbf{(c)} Same as (a)/(b) but for $x=0.78$. Inset shows the cut geometry again but shows the FS has changed through the Lifshitz transition at this doping.
\textbf{(d)} Spectrum of $x=0.41$  below $T_c$ for a different experimental geometry. Inset shows where the spectrum cut is taken to maximize the signal of the $d_{xz}$ orbital around $M$. Red dotted line shows where the EDC is taken for (i). Inset schematic shows where the spectrum cut is taken with respect to the FS in an experimental geometry to maximize the signal of the $d_{xy}$ orbital.
\textbf{(e)-(f)} Spectrum of $x=0.53$, and $x=0.78$ below $T_C$ at the spectrum cut indicated by (d). Red dotted line shows where the EDC is taken for (j). Inset in (f) shows the cut geometry again with the FS to emphasize the Lifshitz transition has occurred at this doping level. 
\textbf{(g)} (left)
EDCs at the at momentum indicated by the red and blue lines shown in (a)-(f) representing the $d_{xz}$ (red) and $d_{xy}$ (blue)
bands, respectively. Dotted red/blue lines indicate EDCs for $T<T_C$, and solid lines are the superconducting gap fit to the experimental data. (right) Dotted lines show EDCs taken at $T>T_C$ at the same $k_F$, with solid lines the showing no gap is open above $T_C$. Legend shows colors corresponding to doping levels
of x = 0.41, x = 0.53, x = 0.78. While the gap on the $d_{xz}$ band stays open for all doping levels, the $d_{xy}$ gap closes for doping including and above $x=0.53$, similar to the behavior of the superconducting gap around $\Gamma$.
\textbf{(h)} Gap size for all orbitals around $\Gamma$ and $M$ in the $\Gamma-M$ direction as a function of doping, demonstrating the orbital selective behavior of the superconducting gap. The $d_{yz}$ band around $M$ does not cross $E_F$ before the Lifshitz transition, so there are no markers for $d_{yz}$ $M$ prior to $x=0.78$ doping.
}
\label{fig:mbandssuper}
\end{figure*}

In the superconducting state, we find that the gap structure of the three hole bands around $\Gamma$ of optimally doped (x=0.41) Ba$_{1-x}$K$_{x}$Fe$_{2}$As$_{2}$ is consistent with previous reports \cite{CAI20211839, PhysRevB.89.064514,PhysRevLett.105.117003}. We identify a nearly isotropic gap on all three hole bands around $\Gamma$, with the largest gap of 9.4 meV in the $\alpha$ sheet, and the smallest gap of 3.1 meV in the $\gamma$ ($d_{xy}$) sheet (\figurename{\ref{fig:gammabandssuper}}(g),(h)). Unlike previous reports at this doping level \cite{CAI20211839, PhysRevLett.105.117003, Ding_2008}, we are able to detect a gap in the $d_{xy}$ electron pocket around $M$ (\figurename{\ref{fig:mbandssuper}}(a),(g)), which has a similar gap size to the $d_{xy}$ sheet around $\Gamma$.

When the doping is increased to $x=0.53$, which is still below the Lifshitz transition, we see there is no longer a gap in the symmetrized energy distribution curve (EDC) for the $d_{xy}$ band around both $\Gamma$ and $M$(\figurename{\ref{fig:gammabandssuper}}(g), \figurename{\ref{fig:mbandssuper}}(g)). This gap closure for the $d_{xy}$ band has also been reported in another ARPES study around $\Gamma$ \cite{PhysRevB.86.165117}, although the superconducting gap magnitudes of the other bands measured are smaller than our measured results as well as other literature \cite{CAI20211839, PhysRevB.88.220508, PhysRevB.89.064514, PhysRevLett.105.117003}. We cannot exclude that there may exist a very small gap in the $d_{xy}$ band that is beyond our experimental resolution, as specific heat studies suggest the presence of very small gaps \cite{PhysRevB.94.205113}. We also see larger anisotropy in the superconducting gap of the $\beta$ sheet from $\Gamma-M$ to $\Gamma-X$ at $x=0.53$ when compared to the doping of $x=0.41$ (\figurename{\ref{fig:gammabandssuper}}(h)).

We further study $x=0.78$ doping, which is beyond the Lifshitz transition, such that the $d_{yz}$ hole band crosses $E_F$. We see the gap stays closed in the $d_{xy}$ band, while the gaps in the other bands around $\Gamma$ shrink but stay finite at around 2.5 meV. This is different from a recent study of KFe$_{2}$As$_{2}$, which finds the $d_{yz}$ band develops nodes in $\Gamma-X$ \cite{Wu2024}. This discrepancy could indicate that the nodes in KFe$_{2}$As$_{2}$ are accidental, which is supported by complementary magnetic penetration depth and specific heat measurements \cite{Wilcox2022, PhysRevB.94.205113}. Around $M$ we do not see a gap on either the $d_{xy}$ or $d_{yz}$ band. This result also differs from reports at higher doping, where a finite gap on the $d_{xy}$ band at $M$ was detected \cite{PhysRevB.88.220508, Wu2024}. While we have not measured the superconducting gap at these doping levels, we note that we observed a strong photon energy dependent suppression of spectral weight due to matrix element effects, of the $d_{xy}$ band, which could affect the determination of the superconducting gap. Some photon energies give a weaker $d_{xy}$ band signal, and thus may give the appearance of a finite superconducting gap while such a gap is absent atphoton energies with stronger $d_{xy}$ band signal (Supplementary Fig. S6, S8). Additionally, past the Lifshitz transition, the orbital character mixing due to hybridization between the $d_{xy}$ and $d_{yz}$ bands may affect the gap magnitude and gap anisotropy. The residual $d_{xz}$ electron band displays a gap that opens in the superconducting state and closes in the normal state. (\figurename{\ref{fig:mbandssuper}}(g)). The superconducting gap measurements are summarized in \figurename{\ref{fig:gammabandssuper}} and \figurename{\ref{fig:mbandssuper}}.\par
\begin{figure*}
\includegraphics[width=\textwidth]{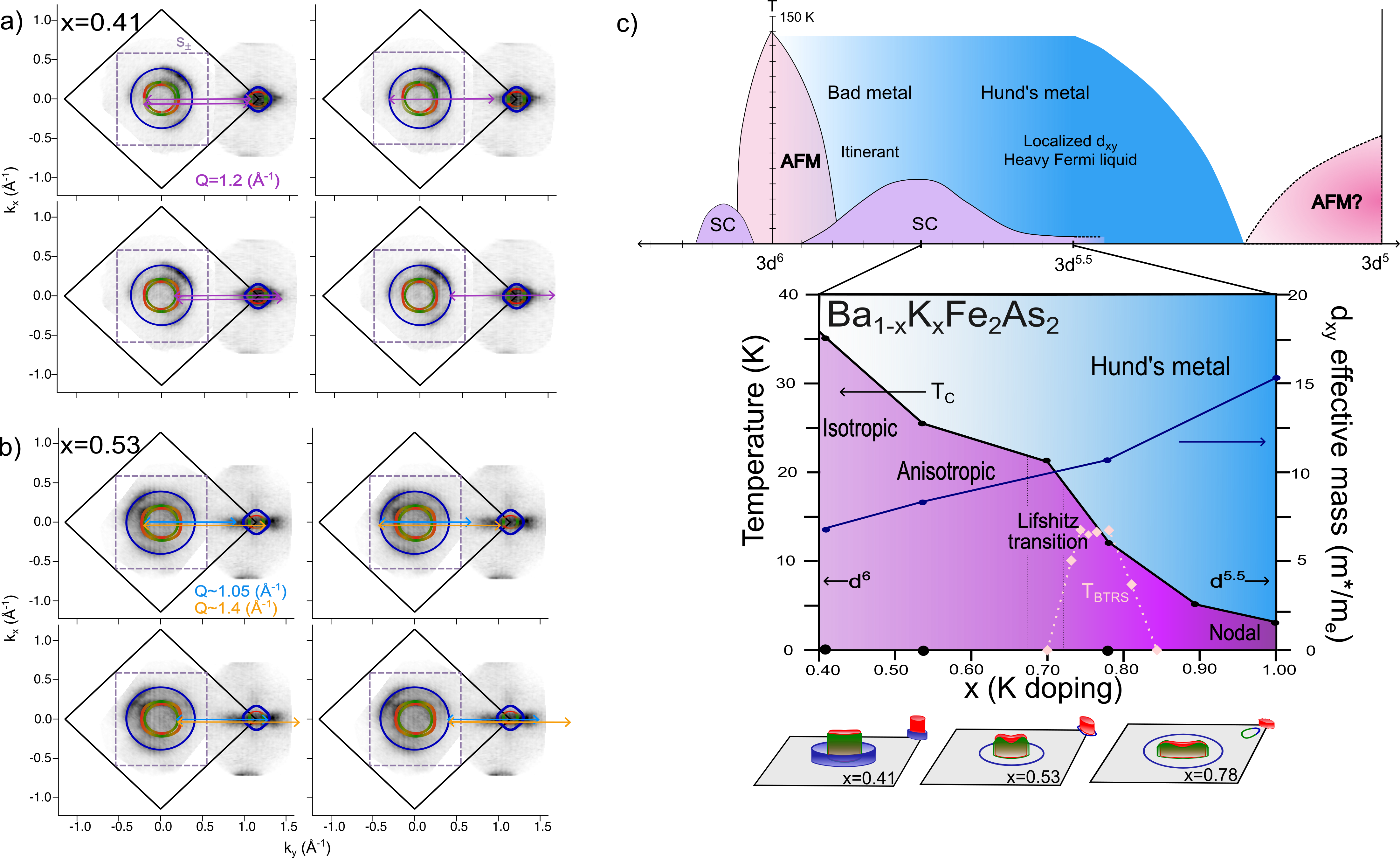}
\caption{\label{fig:scatteringphasediagram}\textbf{Scattering properties and phase diagram of 122 family.} \textbf{(a)} 
Experimental FS for $x=0.41$, with the experimentally determined orbital character overlayed. The solid black line shows the Brillouin Zone, and the dotted purple line shows $s_{\pm}$ nodal lines . (Top left) The magnetic excitation $Q$ vector taken from \cite{PhysRevLett.107.177003} is shown placed starting at the largest gaps on the left side of the hole $\Gamma$ $\alpha, \beta$ bands. (Bottom left) Same as above, but placed on the right side of the $\alpha, \beta$ pockets. (Top/bottom right) Same as left but for the $\gamma$ $d_{xy}$ pocket (left/right) hand side FS. The scattering vector connects the $\alpha, \beta$ pockets with both electron pockets around $M$ fairly well, but is poorly connects the $\gamma$ $d_{xy}$ FS. 
\textbf{(b)} 
Same as \textbf{(a)} but for doping of $x=0.53$, demonstrating scattering does not necessarily favor the $d_{xz/yz}$ pockets. \textbf{(c)} (Top) Cartoon global phase diagram of BaFe$_{2}$As$_{2}$ with electron and hole doping. Note that relative temperatures are not exactly to scale. (Bottom) Zoomed in phase diagram for a region of hole-doped Ba$_{1-x}$K$_{x}$Fe$_{2}$As$_{2}$. Black dots correspond to left axis depicting the superconducting transition temperature. Dark blue dots correspond to $d_{xy}$ band effective mass on the right side axis. The bottom includes inset depictions of the gap structure for doping levels of $x=0.41$, $x=0.53$, $x=0.78$.}
\end{figure*}

What has emerged is a superconducting gap that is highly orbital dependent, with the $d_{xy}$ gap decreasing to diminishing value at $x=0.53$ and beyond, while the gaps of other bands persist up to $x=0.78$ (and beyond) with increasing anisotropy. In the recent study on KFe$_2$As$_2$, the authors  speculated that the suppressed gap on the $d_{xy}$ band may be attributed to the $d_{xy}$ FS’ proximity to the $s_{\pm}$ nodal lines \cite{Wu2024}. The $s_{\pm}$ gap symmetry is defined as $|\cos(k_x) +\sin(k_x)|$. In contrast, we show in \figurename{\ref{fig:scatteringphasediagram}}(a,b) that neither the blue $d_{xy}$ hole nor electron FS is significantly closer to the $s_{\pm}$ nodal lines, here represented by the dotted purple line, at $x=0.53$, where we measure a gap of zero. This calls into question the mechanism by which the gap goes to zero on the $d_{xy}$ band, which we will discuss in more detail later in the article. 

\section{\label{sec:Normalsstatemeasurements}Normal State Measurements}
The change in gap symmetry at $x=0.53$ is at first surprising given that the system has not gone through a Lifshitz transition yet, and therefore one might expect the same bands to participate in pairing. To better understand the superconducting properties, we next investigate the normal state behavior. We utilize DFT that takes into account the variation of lattice constants to account for screening effects and compare these calculations to the experimental results. The DFT band renormalization results of all doping levels are shown in Supplementary Fig. S12. We find there is a clear trend of increasing $d_{xy}$ band renormalization with increasing doping, as compared to the other orbitals, as shown in Table \ref{table:table1}. This is consistent with previous reports of orbital selective correlations due to the interplay between Coulomb interactions and Hund’s coupling \cite{Yi2015, Yi2017}. 

\begin{table*}
\caption{
\textbf{DFT band renormalization values} Table of the band renormalization factors determined by fitting experimental data in the vicinity of $E_F$. The $d_{xz}$ band is omitted due to the large fitting error bars associated with the narrow bandwidth, as well as the uncertainty in orbital determination past the Lifshitz transition. Full fitting results are shown in Supplementary Table S1. These results show the increase in $d_{xy}$ band renormalization as doping is increased, while the other orbitals do not show the same clear trend.}
\begin{ruledtabular}
\begin{tabular}{ddddddd}
\multicolumn{1}{c}{\textrm{Doping level}}&
\multicolumn{1}{c}{\textrm{$d_{xz}$ $\Gamma$}}&
\multicolumn{1}{c}{\textrm{$d_{yz}$ $\Gamma$}}&
\multicolumn{1}{c}{\textrm{$d_{xy}$ $\Gamma$}}&
\multicolumn{1}{c}{\textrm{$d_{yz}$ M}}&
\multicolumn{1}{c}{\textrm{$d_{xy}$ M}}\\
\hline
x=0.41&3.84&2.85&4.54&3.84&4.76\\
x=0.53&  3.70&  2.85&  5.26&3.70& 5.26\\
x=0.78&  3.33&  2.85&  6.25& 2.77& 7.14\\
x=1.00&  4.0&  4.50&  8.33&2.70&  11.11\\
\end{tabular}
\end{ruledtabular}
\label{table:table1}
\end{table*}

\begin{figure*}
\includegraphics[width=\textwidth]{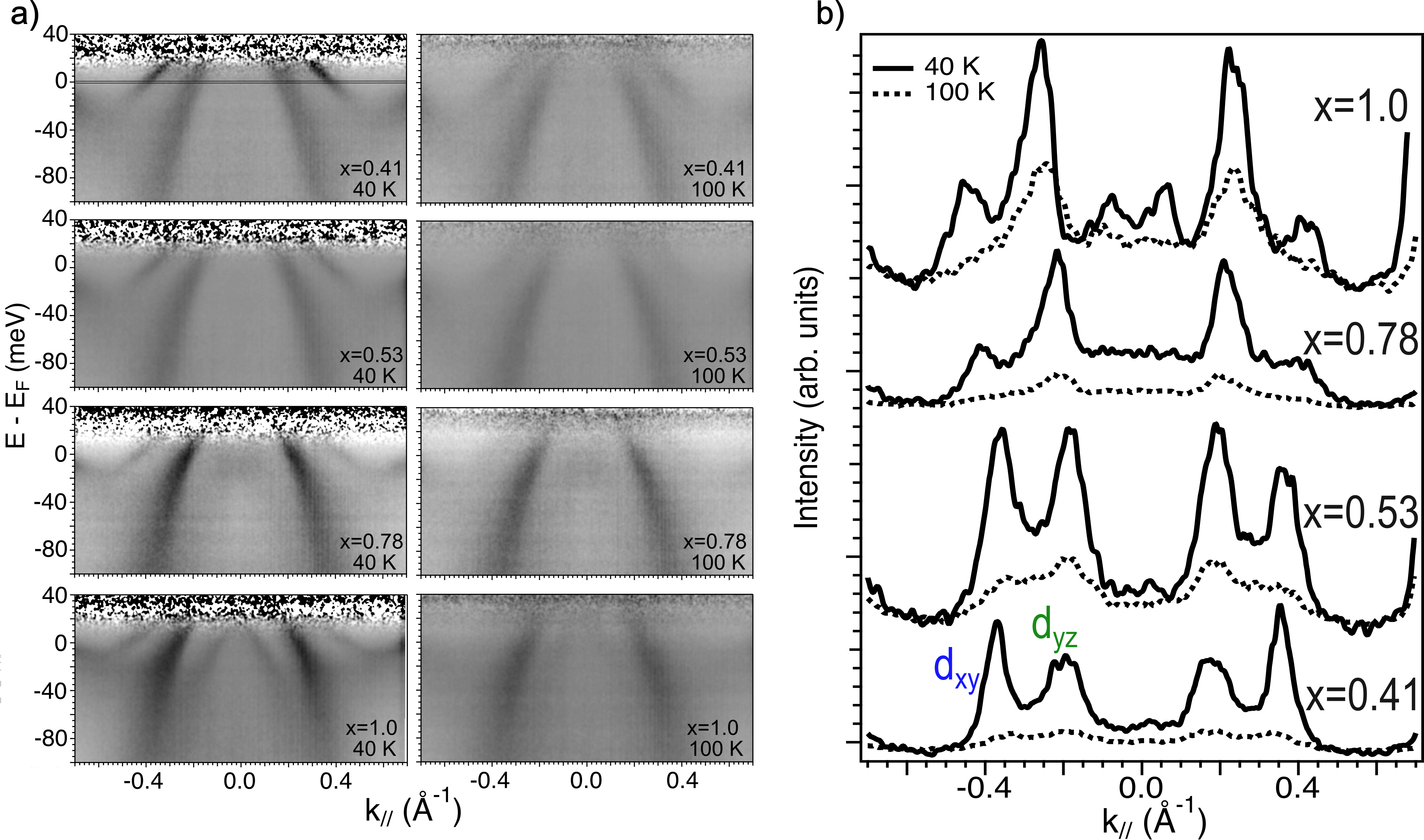}
\caption{\textbf{Temperature dependent incoherence and spectral weight transfer.}
\textbf{(a)} Background subtracted spectra at 40 K (left) and 100 K (right) for dopings of $x=0.41$ (top), $x=0.53$, $x=0.78$, $x=1.0$ (bottom) in the $\Gamma-M$ direction. The black lines in the first spectrum show the representative energy range around $E_F$ where the MDC is taken.
\textbf{(b)} MDCs around $E_F$ at 40 K (solid) and 100 K (dotted) for all doping levels depicting the disappearing $d_{xy}$ hump as a function of temperature, most visibly for doping levels of $x=0.78$, and $x=1.0$, as well as the transfer of intensity from the $d_{xy}$ peak to the $d_{yz}$ with increasing doping. 
}
\label{fig:temperatureincoher}
\end{figure*}

We perform a temperature dependence study for doping levels of $x=0.41$, $x=0.53$, $x=0.78$, and $x=1.0$ and confirm that the $d_{xy}$ band becomes incoherent faster than $d_{xz}/d_{yz}$ as demonstrated in \figurename{\ref{fig:temperatureincoher}}. The spectra shown, and corresponding momentum distribution curves (MDCs) have background signal subtracted and are divided by the Fermi-Dirac function to see above $E_F$ (Supplementary Fig. S11). Both \figurename{\ref{fig:temperatureincoher}}(a), (b)
 indicate the $d_{xy}$ band has weaker intensity with increasing doping and temperature, while the $d_{yz}$ peaks become sharper with doping, and mostly retain their intensity with increased temperature. This spectral weight redistribution of the $d_{xy}$ orbital to the $d_{xz}$/$d_{yz}$ orbitals around $E_F$ has been theoretically predicted for increasing Hund's coupling \cite{hundsspectralweightsuperPhysRevLett.125.177001}. At 100 K, the $d_{xy}$ band peak begins to vanish as doping is increased, with diminished signal by $x=1.0$. The band renormalization and temperature dependence results clearly demonstrate that as the system moves towards n=5.5, the $d_{xy}$ orbital approaches half filling and localization before the $d_{xz/yz}$ orbitals, indicative of orbital selective localization-delocalization behavior.

\section{\label{sec:Discussion}Discussion}
Our results point to a strongly orbital dependent system in both the superconducting and normal state. We now attempt to understand this behavior as a whole. We first investigate if the doping evolution of the $d_{xy}$ band superconducting gap can be trivially explained by poorer nesting behavior within the weak-coupling itinerant spin fluctuation model. Shown in \figurename{\ref{fig:scatteringphasediagram}}(a) is the scattering vector taken from \cite{PhysRevLett.107.177003} for optimal doping overlaid on the experimentally determined ARPES FS. We show that the scattering vector connects the $d_{xz/yz}$ hole bands around $\Gamma$ to both $d_{xz/yz}$ and $d_{xy}$ electron pockets around $M$ fairly well, meanwhile the $\Gamma$ $d_{xy}$ pocket is less well connected to the $M$ pockets. However, the gap size of both $d_{xy}$ bands is very similar in size, and much smaller than the $d_{xz/yz}$ band gaps. This suggests that intra-orbital scattering may be the dominant pairing interaction over inter-orbital, as has been theoretically suggested \cite{Hirschfeld_2011, twodome, rongyuorbitalselectivesuper}. Also in \figurename{\ref{fig:scatteringphasediagram}}(b), we show the scattering vector of $x=0.5$ \cite{PhysRevLett.107.177003} overlaid for our $x=0.53$ FS. Here, it actually appears the $d_{xy}$ bands at both $\Gamma$ and $M$ are better connected as compared to optimal doping, and yet the $d_{xy}$ superconducting gap is zero for $x=0.53$ around both $\Gamma$ and $M$. Additionally, the scattering of the $d_{xz/yz}$ bands appears more poorly connected than at optimal doping, while superconductivity persists for these orbitals. This leads us to speculate that the Fermiology may play a secondary role for the superconducting gap structure.

This doping evolution of the orbital-selective superconductivity may be explained by the increased Hund’s coupling influence and subsequent strong correlation effects. A previous study intending to uncover the origin of anisotropic superconducting gaps in other iron-based superconductors showed that within an itinerant spin-fluctuation model, it may be necessary to include the quasiparticle spectral weights to the susceptibility to reproduce the experimentally determined anisotropic gap structures \cite{orbitalselectiveFeSesuperdoi:10.1126/science.aal1575, orbitalselective_weakcouplign}. This study demonstrated that reduction of quasiparticle weight leads to suppression of pair scattering. Given that the $d_{xy}$ band exhibited clear suppression of spectral weight with increasing hole doping, this may explain the diminishing $d_{xy}$ gap size. Additionally, another study, which took into account dynamical effects, showed that increasing Hund’s coupling not only led to large orbital dependent anisotropy in the superconducting gap sizes, with $d_{xy}$ being the smallest, but also led to superconductivity that persists for more hole-doping values \cite{hundsspectralweightsuperPhysRevLett.125.177001}. This is consistent with the asymmetry between the electron- and hole-doped sides of the phase diagram, where superconductivity extends to a much wider hole-doping range than the electron-doped side.

Recent work \cite{heavyfermioncsfe2as2PhysRevLett.134.076504, PhysRevB.107.115144} has extended hole doping in Ba$_{1-x}$K$_{x}$Fe$_{2}$As$_{2}$ towards Cs(Fe$_{1-x}$Cr$_{x}$)As$_2$, driving the system closer to half filling ($n=5$). In this regime, the material exhibits heavy fermion behavior and signatures of an emerging antiferromagnetic order, suggesting that Ba$_{1-x}$K$_{x}$Fe$_{2}$As$_{2}$ lies near the boundary between itinerant and localized electron behavior. Traditionally, the iron pnictide phase diagram is understood as doping away from the $3d^{6}$ configuration (\figurename{\ref{fig:scatteringphasediagram}}(c)). In this framework, superconductivity emerges from a bad metal state \cite{Werner2012, Fernandes2022}. However, if one instead considers the progression starting at $3d^{5}$, the phase diagram begins to resemble those of other unconventional superconductors like the cuprates, where superconductivity appears after doping away from a strongly correlated insulating state \cite{doi:10.1126/science.235.4793.1196, sobotareview}. If the true parent state is taken to be the n=5 configuration, it has been speculated that this would explain the asymmetry between the electron- and hole-doped sides of the phase diagram and would be consistent with the cuprate phase diagram \cite{mottparentphase}. In this context, the role of the $d_{xy}$ electrons becomes particularly intriguing: they may serve as a bridge between the localized picture and the itinerant iron-based superconductors, acting as more strongly correlated, localized electrons in the background of the itinerant $d_{xz}/d_{yz}$ electrons that are free to pair.

BTRS been shown to exist in this system for a number of doping levels in proximity of $x=0.78$ \cite{Grinenko2020, grinenko2021state, shipulin2023calorimetric}, and our measured superconducting gap structure may provide possible insight into this phenomenon. Minimalistic theoretical models suggest that the presence of three superconducting gaps of similar magnitude, under certain conditions, allows formation of  BTRS through frustration of the superconducting phases \cite{Grinenko2020, grinenko2021state}. Our results have shown that at $x=0.78$ there are three distinct superconducting gaps of similar magnitude, with other gaps suppressed, thus supporting this theoretical framework. However, analysis of fluctuations above the superconducting phase transition in \cite{grinenko2021state} also hinted that minimalistic three-gap models are insufficient to account for available phenomenology, while correlation effects appear to be essential and must be understood. Furthermore, the demonstrated richness of the system may allow for various mechanisms to describe the BTRS superconducting state. For example, given the superconducting pairing on the $d_{xy}$ band becomes vanishingly weak, this could allow a quasi-local moment to form on this orbital. While the presence of local moment-related magnetism in this doping range was challenged from recent NMR and $\mu$sR studies \cite{nmrpaper}, fluctuating local moments at higher frequencies could exist. In this case, as a result of the fluctuating local moment, virtual hopping of Cooper pairs from the $d_{xz}$ to the $d_{yz}$ band through the intermediate $d_{xy}$ orbital could acquire a complex amplitude, thereby breaking time-reversal symmetry. At higher doping levels, the moments would be screened by the more itinerant electrons due to the Kondo effect, such that BTRS would go away. The precise model of BTRS in Ba$_{1-x}$K$_{x}$Fe$_{2}$As$_{2}$ remains an interesting subject for further investigations.\par
Our work has demonstrated the importance of Hund’s coupling to understand both the normal and superconducting state of Ba$_{1-x}$K$_{x}$Fe$_{2}$As$_{2}$. The $d_{xy}$ orbital has been shown to control emergent properties of the iron chalcogenides \cite{Huang2022, Morfoot2023, nontrivialtopologyPhysRevLett.132.136504}, which tend to show the largest effective masses and band renormalization \cite{Yi2015, Yin2011} in the iron-based superconductors. Our results in the pnictides demonstrate the breadth of systems where Hund’s coupling plays an integral role in the emergent physics. These results also hold significant consequences for other multiorbital superconducting systems, such as the nickelates \cite{nickelateKang2023, nickelatehunds}. 

\section*{Acknowledgements}
We thank Yingfei Li for fruitful discussions. This material is based upon work supported by the National Science Foundation Graduate Research Fellowship under Grant No.2021310977 (E.C.). Preliminary data was taken at Diamond Light Source I05 beamline and BESSY One-Cube beamline. The work at Stanford University and Stanford Institute for Materials and Energy Sciences is supported by the Department of Energy, Office of Basic Energy Sciences, Division of Materials Sciences and Engineering, under contract DE-AC02-76SF00515 (E.C., R.Z., T.P.D., and Z.-X.S.). Use of the Stanford Synchrotron Radiation Lightsource, SLAC National Accelerator Laboratory, is supported by the US Department of Energy, Office of Science, Office of Basic Energy Sciences, under contract no. DEAC02-76SF00515 (E.C., M.H., DH.L., and Z.-X.S.). Computational work was performed on the Sherlock cluster at Stanford University and on resources of the National Energy Research Scientific Computing Center (NERSC), a Department of Energy Office of Science User Facility, using NERSC award BES-ERCAP0031425 (R.Z.,T.P.D.). E.B. was supported by the Swedish Research Council Grant   2022-04763, Olle Engkvists Stiftelse, a project grant from Knut och Alice Wallenbergs Stiftelse,  and partially by the Wallenberg Initiative Materials Science
for Sustainability (WISE) funded by the Knut and Alice Wallenberg Foundation. V.G. is supported by the NSFC grant 12374139.


\bibliography{apssamp}

\appendix

\end{document}